\documentclass[prd, aps, nofootinbib,preprintnumbers, showpacs,superscriptaddress,twocolumn]{revtex4}
\usepackage{latexsym}
\usepackage{amssymb}
\usepackage{amsfonts}
\usepackage{amsmath}
\usepackage{bm}
\usepackage[dvips]{graphicx}
\usepackage[usenames,dvipsnames]{color}
\usepackage{subfigure}
\usepackage{times}
\usepackage{units}
\usepackage{hyperref}

\usepackage[utf8x]{inputenc}
\usepackage{amssymb,amsmath}
\usepackage{graphicx}
\usepackage[squaren]{SIunits}
\graphicspath{{./plots/}}

\newcommand{\Fig}[1]{Fig.~\ref{#1}}

\newcommand{\be}{\begin{equation}}
\newcommand{\ee}{\end{equation}}
\newcommand{\bea}{\begin{eqnarray}}
\newcommand{\eea}{\end{eqnarray}}
\newcommand{\beqn}{\begin{eqnarray*}}
\newcommand{\eeqn}{\end{eqnarray*}}
\newcommand{\ba}{\begin{align}}
\newcommand{\ea}{\end{align}}

\newcommand{\Msun}{\,M_{\odot}}

\newcommand{\cf}{\textit{cf.}~}
\newcommand{\ie}{\textit{i.e.,}~}
\newcommand{\eg}{\textit{e.g.,}~}

\begin{document}


\title{Constraint damping of the conformal and covariant formulation of the
   Z4 system \\
in simulations of binary neutron stars}


\author{Daniela \surname{Alic}}
\affiliation{Max-Planck-Institut f\"ur Gravitationsphysik,
  Albert-Einstein-Institut, Potsdam 14476, Germany}

\author{Wolfgang \surname{Kastaun}}
\affiliation{Max-Planck-Institut f\"ur Gravitationsphysik,
  Albert-Einstein-Institut, Potsdam 14476, Germany}

\author{Luciano \surname{Rezzolla}} 
\affiliation{Max-Planck-Institut f\"ur Gravitationsphysik,
  Albert-Einstein-Institut, Potsdam 14476, Germany}
\affiliation{Institute for Theoretical Physics, Max von Laue Strasse 1,
  60438 Frankfurt am Main, Germany}


\begin{abstract}
Following previous work in vacuum spacetimes, we investigate the
constraint-damping properties in the presence of matter of the recently
developed traceless, conformal and covariant Z4 (CCZ4) formulation of the
Einstein equations. First, we evolve an isolated neutron star with an
ideal gas equation of state and subject to a constraint-violating
perturbation. We compare the evolution of the constraints using the CCZ4
and Baumgarte-Shibata-Shapiro-Nakamura-Oohara-Kojima (BSSNOK) systems. Second, we study the collapse of an unstable
spherical star to a black hole. Finally, we evolve binary neutron star
systems over several orbits until the merger, the formation of a black
hole, and up to the ringdown. We show that the CCZ4 formulation is stable
in the presence of matter and that the constraint violations are 1 or
more orders of magnitude smaller than for the BSSNOK
formulation. Furthermore, by comparing the CCZ4 and the BSSNOK
formulations also for neutron star binaries with large initial constraint
violations, we investigate their influence on the errors on physical
quantities. We also give a new, simple and robust prescription for the
damping parameter that removes the instabilities found when using the
fully covariant version of CCZ4 in the evolution of black holes. Overall,
we find that at essentially the same computational costs the CCZ4
formulation provides solutions that are stable and with a considerably
smaller violation of the Hamiltonian constraint than the BSSNOK
formulation. We also find that the performance of the CCZ4 formulation is
very similar to another conformal and traceless, but noncovariant
formulation of the Z4 system, \ie the Z4c formulation.
\end{abstract}


\maketitle


\section{Introduction}
\label{sec:intro}

Recent developments in numerical relativity allow for the simulation of
binary compact objects, \eg binary neutron star (NS) systems, which are
among the strongest sources of gravitational waves. Given that less than
one decade ago, even the vacuum binary black hole (BH) problem was still
a challenge for numerical evolutions of the Einstein equations, the
progress is indeed remarkable. Part of this rapid progress is surely due
to a better understanding of the mathematical properties of the different
formulations of the Einstein equations and the choice of suitable
gauge conditions. The most widely used formulation of the Einstein
equations in three-dimensional numerical simulations is given by a
conformal and traceless formulation of the ADM (Arnowitt-Deser-Misner)
3+1 decomposition of the Einstein equations~\cite{Arnowitt62,York79}, and
is also known as the 
BSSNOK (or BSSN) formulation~\cite{Nakamura87,
  Shibata95, Baumgarte99}.

Another conformal and traceless formulation, \ie the 
CCZ4 formulation, was introduced recently in
Ref.~\cite{Alic:2011a} (hereafter paper I), where we discussed in detail
its structure and properties. The CCZ4 formulation shares some important
properties with the BSSNOK formulation. It allows for stable evolutions in conjunction
with robust and singularity-avoiding gauges, thus eliminating the need
for excision techniques, and with simple Sommerfeld radiative boundary
conditions. The main advantage of the CCZ4 formulation over the BSSNOK
formulation is its ability to dynamically evolve the constraints
equations and to couple them with a constraint-damping scheme that leads
to a rapid suppression of the violations that are inevitably produced in
numerical simulations.

In paper I, we validated the properties of the CCZ4 system in evolutions
of single and binary BH systems. The constraint-propagating and damping
properties of the system turned out be very useful in reducing the
violations produced in long-term simulations and led to constraint
violations that were, on average, 1 order of magnitude smaller than
those with the BSSNOK system. In this work, we investigate the behavior
of the CCZ4 system also in evolutions of nonvacuum spacetimes, in
particular those of single and binary NSs. In paper I, the simulations were
performed with both a multipatch coordinate system, where the spherical
outer boundary was causally disconnected, and Cartesian grids plus
radiative boundary conditions. In the latter system, we noticed reflections from
the outer boundary which led to an increase in the values of the
constraints. However, these violations were rapidly damped, and there was
no sign of instabilities produced in relation with the outer boundary. In
this work, we focus on the latter case, which is used more often in
simulations of neutron star mergers.

We recall that the CCZ4 system contains three constant parameters, one of
which ($\kappa_3$, see below) determines if the system is actually
covariant or not (throughout this paper we refer to both versions as
CCZ4). It is obviously advantageous if the propagation equations for the
constraints are covariant, since it allows one to generalize results
found in a particular coordinate system. For noncovariant formulations,
on the other hand, the behavior of the constraint evolution could in
principle be radically altered when going, for example, from Cartesian to
spherical coordinates. This is relevant in practice when comparing
one-dimensional and three-dimensional evolutions or when using multipatch
coordinate systems. While both covariant and noncovariant formulations
were studied in paper I, we also found there that exponentially growing
modes appear for the fully covariant system. An important point of this
paper is the exploration of the behavior of the covariant and
noncovariant versions of the CCZ4 system in spacetimes containing
matter. We will show that the aforementioned instabilities occur only
when a BH is present in the spacetime, while simulations of NSs do not
show such instabilities. Furthermore, we will demonstrate that even when
a BH appears in the solution, \eg in the collapse of a NS, or after the
merger of two NSs, a modification of the damping terms in the fully
covariant CCZ4 system results in stable evolutions. Finally, we will
also confirm that in all comparisons between the covariant and the
noncovariant versions of the CCZ4 system, the differences between the
results are very small.

A different conformal version of the noncovariant Z4 system has been
presented in Refs.~\cite{Bernuzzi:2009ex, Ruiz:2010qj, Weyhausen:2011cg,
  Cao:2012, Hilditch2012}. This system, called Z4c, removes source terms
in order to bring the evolution equations into a form which is closer
to the
BSSNOK system. Recent numerical results obtained with Z4c in
three-dimensional simulations of compact objects~\cite{Hilditch2012}
report a reduction in the Hamiltonian constraint violation between 1
and 3 orders of magnitude with respect to BSSNOK for nonvacuum
spacetimes, and between 1 and 2 orders of magnitude for black hole
spacetimes. Performing a similar comparison of the Z4c and BSSNOK
systems, we could measure a reduction of constraint violations of only
1 or 2 orders of magnitude, probably because we lack the improvement
associated with Z4c, that comes from the use of constraint-preserving boundary
conditions of the type described in Ref.~\cite{Ruiz:2010qj}. The direct
comparison between the CCZ4 and Z4c systems shows that the CCZ4 leads to
similarly low constraint violations and allows long-term stable evolution
when coupled with standard radiative boundary conditions. Analytically,
there are no features intrinsic to the Z4c system which suggest
improvement over CCZ4.

The plan of the paper is as follows: Section~\ref{sec:num_meth} presents
an overview of the CCZ4 system of the Einstein equations coupled with the
hydrodynamic equations, as well as the numerical infrastructure of our
simulations. Section~\ref{sec:results} is dedicated to numerical results
obtained with the CCZ4 system in evolutions of a stable, isolated NS,
of the
collapse of an unstable NS to a BH, and of binary NS systems, both with
constraint-satisfying and constraint-violating initial data. In the
first four subsections, we compare results obtained with the BSSN and
CCZ4 formulations of the Einstein equations, while the last two
subsections are dedicated to a comparison with the Z4c system and between
the fully covariant and the noncovariant versions of the CCZ4 system. We
conclude with the summary and discussions in Sec.~\ref{sec:summary}.

Throughout this paper, we use a metric signature of $(-,+,+,+)$ and units
in which $c=G=M_\odot =1$, unless noted otherwise. Greek indices are
taken to run from 0 to 3, Latin indices from 1 to 3, and we adopt the
standard convention for summation over repeated indices.

\section{Formulation and Numerical Methods}
\label{sec:num_meth}

\subsection{The CCZ4 system}
\label{sec:CCZ4}

The CCZ4 system is a conformal and covariant formulation of the Einstein
equations. It is based on a conformal transformation of the Z4 system
(see paper I), which converts the original Hamiltonian and momentum ADM
constraints into evolution equations for a four-vector $Z_{\mu}$. This
amounts to introducing two additional evolution variables, namely the
projection $\Theta$ along the normal direction of the four-vector
$Z_{\mu}$ and its spatial component $Z_i$. The system can be supplemented
with damping terms~\cite{Gundlach2005:constraint-damping}, which ensure
exponential damping of the constraint violations in numerical evolutions.

The steps required to convert the original Z4 system into a conformal
version were presented in our previous paper I. For completeness, we
summarize here the main ideas behind the Z4 formulation and point to Refs.~\cite{Bona:2003qn, Alic:2009} for more details about its properties.
The CCZ4 system is given by the following set of evolution equations:
\begin{widetext}
\begin{eqnarray}
\partial_t\tilde\gamma_{ij} &=& - 2\alpha \tilde A^{^{\rm TF}}_{ij}
+ 2\tilde\gamma_{k(i}\partial_{j)}~\beta^k
- \frac{2}{3}\tilde\gamma_{ij}\partial_k~\beta^k
+\beta^k \partial_k \tilde\gamma_{ij} \,,  \label{gamma_eq}\\
\partial_t \tilde A_{ij} &=& \phi^2  \left[-\nabla_i \nabla_j \alpha
+ \alpha \left(R_{ij} + \nabla_i Z_j + \nabla_j Z_i - 8 \pi S_{ij}\right)\right]^{\rm TF}
+ \alpha \tilde A_{ij}\left(K- 2\Theta\right) \nonumber \\
&&
- 2\alpha \tilde A_{il}\tilde A^l_j 
+ 2\tilde A_{k(i} \partial_{j)} \beta^k
-\frac{2}{3}\tilde A_{ij} \partial_k \beta^k + \beta^k \partial_k \tilde A_{ij}  \,,  \label{A_eq} \\
\partial_t\phi &=& \frac{1}{3} \alpha \phi K
- \frac{1}{3} \phi \partial_k \beta^k + \beta^k \partial_k \phi \,, \\
\label{eq:dt_K}
\partial_t K &=& - \nabla^i \nabla_i \alpha + \alpha \left(R + 2
  \nabla_iZ^i + K^2 -2 \Theta K \right)
+ \beta^j \partial_j K - 3 \alpha \kappa_1 \left(1 +
\kappa_2\right) \Theta
+ 4 \pi \alpha \left(S - 3 \tau\right) \,, \label{K_eq}\\
\label{eq:dt_Theta}
\partial_t \Theta &=& \frac{1}{2} \alpha \left(R + 2 \nabla_i Z^i - \tilde A_{ij} \tilde
A^{ij} + \frac{2}{3} K^2 \,\textcolor{blue}{\boxed{- 2 \Theta K}} \right) \,\textcolor{blue}{\boxed{- Z^i
\partial_i \alpha}} + \beta^k \partial_k \Theta
- \alpha \kappa_1 \left(2 + \kappa_2\right) \Theta - 8\pi \alpha\,\tau\,, \\
\partial_t \hat\Gamma^i &=& 2\alpha \left(\tilde\Gamma^i_{jk} \tilde A^{jk}
- 3 \tilde A^{ij} \frac{\partial_j \phi}{\phi} - \frac{2}{3}
\tilde\gamma^{ij} \partial_j K \right)
+2\tilde\gamma^{ki}\left(\alpha \partial_k \Theta \,\textcolor{blue}{\boxed{- \Theta
\partial_k \alpha
- \frac{2}{3} \alpha K Z_k}}\right) -  2\tilde A^{ij} \partial_j
\alpha \nonumber\\
&& + \tilde\gamma^{kl} \partial_k \partial_l \beta^i
+ \frac{1}{3}\tilde\gamma^{ik}\partial_k\partial_l \beta^l
+ \frac{2}{3} \tilde\Gamma^i \partial_k \beta^k -
\tilde\Gamma^k \partial_k \beta^i 
\,\textcolor{blue}{\boxed{+ 2 \kappa_3 \left(\frac{2}{3} \tilde\gamma^{ij} Z_j \partial_k \beta^k -
 \tilde\gamma^{jk} Z_j \partial_k \beta^i \right)}} \nonumber \\
&&+ \beta^k \partial_k \hat\Gamma^i
- 2 \alpha \kappa_1 \tilde \gamma^{ij} Z_j - 16 \pi \alpha
\tilde\gamma^{ij} S_{j}\,, \label{Gamma_eq}
\end{eqnarray}
\end{widetext}
where ${\tilde{\gamma}}_{ij} = \phi^2 \gamma_{ij} $ is the conformal
metric with unit determinant $\phi = ({\rm det} (\gamma_{ij}))^{-1/6}$,
while the extrinsic curvature $K_{ij}$ is represented by its trace $K
\equiv K_{ij} \gamma^{ij}$ and its trace-free components
\begin{equation}
\label{tracelessK}
{\tilde{A}}_{ij} = \phi^2\;(K_{ij}- \frac{1}{3} K \gamma_{ij})\,.
\end{equation}

The three-dimensional Ricci tensor $R_{ij}$ is split into a part 
$\tilde R^{\phi}_{ij}$ containing conformal terms and another one,
$\tilde R_{ij}$, containing space derivatives of the conformal metric,
defined as
\begin{widetext}
\begin{eqnarray}
\tilde R_{ij} &=& -\frac{1}{2} \tilde \gamma^{lm} \partial_l \partial_m \tilde \gamma_{ij}
  + \tilde \gamma_{k(i} \partial_{j)} \tilde \Gamma^k
  + \tilde \Gamma^k \tilde \Gamma_{(ij)k}
  + \tilde \gamma^{lm} \left[2 {\tilde \Gamma^k}_{l(i}
\tilde \Gamma_{j)km} + {\tilde \Gamma^k}_{im} \tilde \Gamma_{kj\,l}\right]\,, \\
\tilde R^{\phi}_{ij} &=& \frac{1}{\phi^2}\left[\phi \left(\tilde \nabla_i \tilde
\nabla_j \phi + \tilde \gamma_{ij} \tilde \nabla^l \tilde \nabla_l
\phi\right) - 2 \tilde \gamma_{ij} \tilde \nabla^l \phi \tilde \nabla_l
\phi\right]\,.
\end{eqnarray}
\end{widetext}

The following definitions apply for matter-related quantities 
\begin{equation}
\tau \equiv n_{\mu} n_{\nu} T^{\mu\nu}\,, \qquad 
S_i \equiv n_{\nu} T^{\nu}_{~i}\,,  \qquad
S_{ij} \equiv T_{ij}\,,
\end{equation}
and $\Theta$ is the projection of the Z4 four-vector along the normal
direction, $\Theta \equiv n_{\mu} Z^{\mu} = \alpha Z^0$. We note that we
here follow the definition of the normal four-vector suggested in
Ref.~\cite{Bona:2003qn}, \ie $n_{\mu} = (\alpha, 0)$ and $n^{\mu} =
(-1/\alpha, \beta^i/\alpha)$, which is different from the more
traditional one in which $n_{\mu} = (-\alpha, 0)$ and $n^{\mu} =
(1/\alpha, -\beta^i/\alpha)$. These different definitions do not affect
the form of the CCZ4 equations.

The evolution variable $Z_i$ of the Z4 formulation is now
included in the ${\hat \Gamma}^i$ variable of the CCZ4 formulation
\begin{equation}
\label{Gammai}
\hat \Gamma^i \equiv \tilde \Gamma^i + 2 \tilde \gamma^{ij} Z_j\,,
\end{equation}
where 
\begin{equation}\label{localGamma}
\tilde \Gamma^i \equiv \tilde \gamma^{jk} \tilde \Gamma^{i}_{jk} =
\tilde \gamma^{ij} \tilde \gamma^{kl} \partial_l \tilde
\gamma_{jk}\,.
\end{equation}

The numerical simulations presented in this paper use as gauge conditions
the ``$1+\log$'' slicing
\begin{equation}
\label{1plog}
\partial_t \alpha = -2\alpha \left(K-2\Theta\right) + \beta^k\partial_k \alpha \,, 
\end{equation}
and the gamma-driver shift condition
\begin{eqnarray}
\label{gammadriver1}
\partial_t \beta^i &=& f B^i +\beta^k\partial_k\beta^i \,, \\
\label{gammadriver2}
\partial_t B^i &=& \partial_t \hat\Gamma^i - \beta^k \partial_k \hat\Gamma^i
+ \beta^k\partial_k B^i - \eta B^i \,,
\end{eqnarray}
where the gauge parameter $f$ is set to $3/4$. We adopt the
``constrained approach'' from paper I in order to enforce the
constraints of the conformal formulation (trace cleaning).

We also compute the constraint violations introduced by the numerical
evolution in the ADM constraints:
\begin{eqnarray}
\label{eq:coneq_1}
H &=& R - K_{ij} K^{ij} + K^2 - 16 \pi \tau\,, \\
\label{eq:coneq_2}
M_i &=&  \gamma^{jl} (\partial_{l} K_{ij} - \partial_{i} K_{jl} -
\Gamma^{m}_{jl} K_{mi} + \Gamma^{m}_{ji} K_{ml}) 
- 8 \pi S_i\,.
\nonumber\\
\end{eqnarray}
For both the BSSNOK and the CCZ4 systems, we compute the ADM quantities 
from the evolved variables of the two systems. 

We recall that the parameters $\kappa_1$ and $\kappa_2$ control the
damping terms and that all the constraint-related modes are damped when
$\kappa_1 > 0$ and $\kappa_2 >
-1$~\cite{Gundlach2005:constraint-damping}. The third coefficient,
$\kappa_3$, instead affects some quadratic terms in the evolution
for ${\hat \Gamma}^i$ Eq.~\eqref{Gamma_eq} and determines the covariance
of the corresponding CCZ4 system. In particular, a value of $\kappa_3 =1$
corresponds to full covariance. While $\kappa_2$ and $\kappa_3$ are
dimensionless, $\kappa_1^{-1}$ is the damping timescale, which we report
in geometric units with $M_\odot=1$.

We also recall that in paper I we performed numerical experiments with
the covariance parameter $\kappa_3$ and concluded that a choice of
$\kappa_3 = 1$ and $\kappa_1=\rm{const.}$ leads to unstable behavior in
the context of black hole spacetimes. Even though we could not fully
isolate the cause of these numerical instabilities, we found that they
are produced by nonlinear couplings with the damping terms, which are
important for reducing the violations in the constraints. As a result,
all of the tests reported in paper I made use of $\kappa_3 =1/2$, which,
for consistency with paper I, is also the value we will use for the
majority of the tests discussed here. Two important remarks should,
however, be made now, although they will be discussed also later
on. First, in the presence of a nonsingular spacetime with matter, a
fully covariant formulation (\ie with $\kappa_3=1$) does not show any of
the pathologies encountered in paper I with black holes. The pathologies,
however, do appear as soon as a black hole is formed. Second, we have
devised a new prescription for the damping term $\kappa_1$ which couples
it to the lapse function and cures these instabilities also when black
holes appear in the evolution. As a result, independently of whether we
are considering vacuum or nonvacuum spacetimes, our CCZ4 formulation can
now \emph{always} be used in its fully covariant form. A more extended
discussion of this point, with the presentation a series of examples, is
postponed to Sec.~\ref{sec:Cvsc}.

As mentioned in the Introduction, another noncovariant but conformal
formulation of the Z4 system has been proposed recently in
Ref.~\cite{Hilditch2012}, namely the Z4c formulation. The suggestion
behind the Z4c formulation is that of introducing the damping features of
the Z4 formulation with only minimal changes to the BSSNOK system. As we
will comment later on, this strategy is a very reasonable one and leads
to results that are comparable to those of the CCZ4 formulation and whose
main drawback is being noncovariant. While this aspect of the
formulation may be harmless in practice, it makes it difficult to
generalize the properties found in a particular coordinate system,
leaving room for unexpected behavior.

Because the CCZ4 and Z4c formulations differ only in nonprincipal
part terms, we have marked the terms that are missing in the Z4c
formulation with blue boxes within the set of evolution equations
Eqs.~(\ref{gamma_eq})-(\ref{Gamma_eq}). In addition, the Z4c system evolves
the trace of the extrinsic curvature $\hat K$ using an evolution equation
similar to the BSSNOK one~\cite{Hilditch2012}
\begin{eqnarray}
\partial_t \hat K 
&=& - \nabla^i \nabla_i \alpha + \alpha \left(\tilde A_{ij} \tilde
A^{ij} + \frac{1}{3}(\hat K + 2 \Theta)^2 \right) \nonumber\\
& &+ \beta^j \partial_j \hat K + \alpha \kappa_1 \left(1 -
\kappa_2\right) \Theta
+ 4 \pi \alpha \left(S + \tau\right) \, \nonumber \\
\end{eqnarray}
to replace Eq.~(\ref{K_eq}). This variable can be translated in CCZ4 terms
as
\begin{equation}
\hat K  = K^{^{\rm BSSNOK}} = K - 2\, \Theta \,. 
\end{equation}
%

\subsection{The relativistic hydrodynamic equations}
\label{sec:hydro}

We evolve the hydrodynamic equations in flux-conservative form as
\begin{align}
\partial_t \hat{D}
  &= -\partial_k \left[ w^k \hat{D} \right],
\label{rest_mass_evol}\\
\begin{split}
\partial_t \hat{\tau}
  &= -\partial_k \left[ w^k \hat{\tau} + \phi^{-3}\alpha p v^k \right]  \\&\qquad
     +\alpha \hat{S}^{lm} K_{lm} - \hat{S}^k \partial_k \alpha,
\end{split}
\label{energy_density_evol}\\
\begin{split}
\partial_t  \hat{S}_i
  &= -\partial_k \left[ w^k \hat{S}_i + \phi^{-3}\alpha p \delta^k_i \right]  \\&
     \phantom{=}\,\, +\frac{\alpha}{2} \hat{S}^{lm} \partial_i \gamma_{lm} 
     + \hat{S}_k \partial_i \beta^k
     - (\hat{\tau} + \hat{D}) \partial_i \alpha.
\end{split}
\label{momentum_evol}
\end{align}
The evolved variables are the conserved density $\hat{D}$, the conserved
energy density $\hat{\tau}$, and the conserved momentum density
$\hat{S}_i$, whose definition is given by
\begin{align}
\label{Tmunu_decomposition2}
   \hat{D} &\equiv
      \phi^{-3}\rho W\,, \\
   \hat{\tau} &\equiv
      \phi^{-3}\left(\rho h W^2 - p \right) - \hat{D}\,, \\
   \hat{S}_i &\equiv
      \phi^{-3}\rho h W^2 v_i\,.
\end{align}
Above, $\rho$ is the rest mass density, $v^i$ is the 3-velocity, $W$ is
the Lorentz factor, $w^i \equiv \alpha v^i - \beta^i$ is the advection
speed with respect to the coordinates, $p$ is the pressure, $\epsilon$ is
the specific internal energy, and $h = 1+\epsilon + p/\rho$ is the
specific enthalpy. Finally, the projection of the energy-momentum tensor
onto the spatial hypersurfaces of the foliation leads to the
spatial tensor
\begin{align}
\hat{S}_{ij} &\equiv 
\phi^{-3}\left(\rho h W^2 v_{i} v_{j} + \gamma_{ij} p \right)\,.
\end{align}

\subsection{Numerical setup}
\label{sec:numsetup}

To evolve the hydrodynamic equations, we rely on an improved version of
the \texttt{Whisky} code described in Refs.~\cite{Giacomazzo:2007ti,
  Giacomazzo:2009mp, Giacomazzo:2010, Galeazzi2013, Radice2013b}, but
without making use of the ideal MHD part or of the high-order
finite difference operators. The evolution of the spacetime is performed
using our CCZ4 implementation within the \texttt{McLachlan}
code~\cite{Brown2007b}, which is part of the publicly available Einstein
Toolkit infrastructure based on the \texttt{Cactus} computational
framework. For the time integration we are using the method of lines with
an explicit fourth-order Runge-Kutta method. The outer boundary conditions
used are the (Sommerfeld) radiative ones provided by the
\texttt{McLachlan} code, initially developed and well tested for the
BSSNOK system.

We employ adaptive mesh refinement (AMR) for our numerical grid, provided
via the \texttt{Carpet} driver~\cite{Schnetter-etal-03b}. For single-star
simulations, we use a fixed grid hierarchy, while for binary NS runs we
follow the centers of mass of the stars with moving refined boxes. During
the merger, they are replaced by larger nonmoving refinement regions of
the same resolution. Shortly after, one more refinement level with
doubled resolution is activated to better resolve the black hole.

For the solution of the Einstein equations we use finite-difference
spatial differential operators of various orders, although the results
presented in this paper have been obtained using fourth-order-accurate
schemes. The spatial discretization of the hydrodynamic equations, on the
other hand, uses a finite-volume high-resolution shock-capturing scheme
adopting the parabolic reconstruction of the piecewise parabolic method
(PPM)~\cite{Colella84} and the Harten-Lax-van Leer-Einfeldt
(HLLE)~\cite{Harten83} approximate Riemann solver. In contrast to the
original \texttt{Whisky} code, we do not reconstruct the 3-velocities
$v^i$, but rather the quantities $W v^i$ (as done in Ref.~\cite{Galeazzi2013}).
This guarantees that the velocities reconstructed at the cell boundaries
stay subluminal even under the extreme conditions which occur at the
center of a black hole. Using shock tube tests, we verified that this
modification does not affect the treatment of shocks.

Furthermore, we improved the robustness of the conversion algorithm from
evolved to primitive variables, allowing a clear distinction between
physical and unphysical values. The details of this algorithm have
already been described in Ref.~\cite{Galeazzi2013}. For the current work, the
only important aspect of the improvements is the ability to enforce a
fine-grained error policy. The standard policy does not allow any
unphysical values, with the exception of the internal energy falling
below the zero temperature value (zero for the ideal gas equation of
state EOS), in which
case it is reset to that value. This can happen frequently when evolving
cold matter, where the internal energy is exactly at the minimum value
allowed by the given EOS. At densities corresponding
to the surface region of the star, which is a notorious source of errors
in hydrodynamic simulations, we use a more lenient policy, which adjusts
unphysical values of the conserved energy and momentum densities to the
physically meaningful range at a given density. The same applies to a
region around the center of the black hole, which we define as the region
in which $\alpha \leq \alpha_c = 0.1$ and which is always contained
within the apparent horizon. In this region we also limit the Lorentz
factor to be $W \le 3$. This adjustment, together with the aforementioned
modified reconstruction, allows us to evolve a black hole without excision
for the spacetime or for the hydrodynamic variables.

Finally, we added an option to smoothly remove all matter from the center
of a black hole. We did this to investigate how the presence of matter at
the center of the black hole affects the numerical errors in comparison
to an evolution of a vacuum black hole. This needs to be addressed
separately from the spacetime-matter coupling elsewhere, because, although
our gauge is singularity avoiding, the gradients of metric and density
close to the center of a BH still become so large that it is always
severely under-resolved numerically. To remove the matter, we introduce a
term $-q/\tau_d$ to the rhs of the conserved variables $q=(D,\tau,S_i)$.
The baryon mass then evolves according to $\dot{M}_b = - M_b / \tau_d$,
thus leading to an exponential decay.

In simulations that form a black hole, we detect apparent horizons using
the module \texttt{AHFinderDirect}~\cite{Thornburg2003:AH-finding} from
the Einstein toolkit. We compute the mass and spin using the isolated
horizon formalism~\cite{Ashtekar:2004cn,Dreyer02a} implemented in the
\texttt{QuasiLocalMeasures} module. Finally, the
constraints in Eqs.~\eqref{eq:coneq_1}-\eqref{eq:coneq_2} are computed
using a standard fourth-order finite difference method.

\section{Results}
\label{sec:results}

In this section, we investigate the stability of the CCZ4 formulation
when coupled to matter and test the convergence properties of the
code. Furthermore, we compare the behavior of the constraint violations,
first between the CCZ4 and the BSSNOK formulations, then between the CCZ4
and the Z4c formulations, and finally between the covariant and the
noncovariant CCZ4 versions.

\subsection{Isolated neutron star}
\label{sec:isolated}

%
\begin{figure}
  \centering
  \includegraphics[width=0.98\columnwidth]{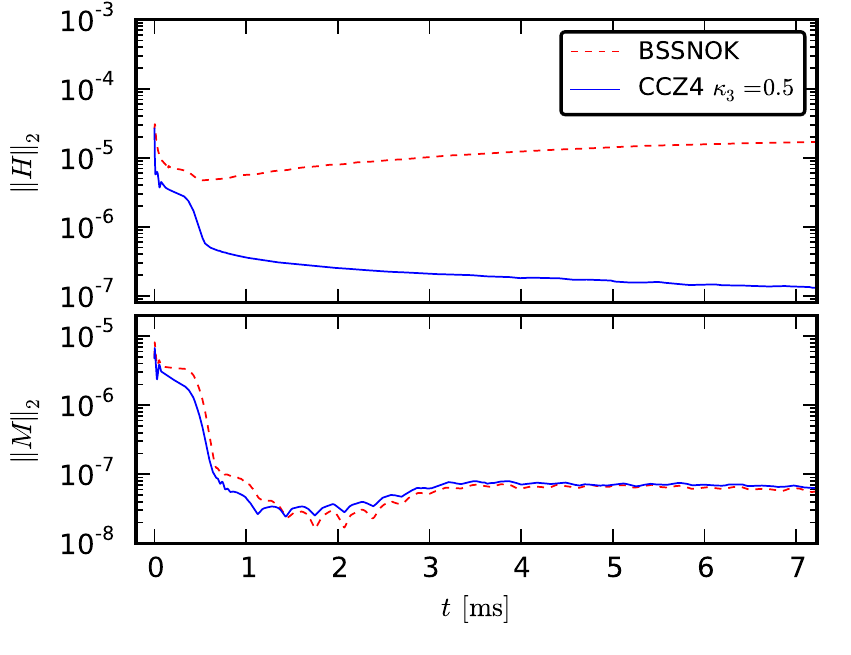}
  \caption{Comparison of the CCZ4 and BSSNOK formulations in the
    evolution of a single and perturbed TOV star. The CCZ4 runs use the
    parameters $\kappa_1=0.02, \kappa_2=0$. Shown are the $L_2$ norms of
    the Hamiltonian (top panel) and the combined momentum (bottom panel)
    constraint violations, taken over the full computational domain.}
  \label{fig:tov_L2constr}
\end{figure}
%

\begin{table*}
\begin{ruledtabular}
\begin{tabular}{lcccccccc}
Simulation (stage) & 
$h_0$ [km]  & 
$h_f$ [km]  & 
Levels    & 
$R_n$ [km] & 
$R_\text{star}$ [km] & 
$R_\text{BH} $ [km] & 
$M [M_\odot]$ &
symmetries \\\hline 
Stable TOV     & 
  $1.1814$ & 
  $0.2953$ & 
  $3~(0)$ & 
  $(94.5, 47.3, 23.6)$ & 
  $14.2$ & 
  -- & 
  $1.40$   &
  octant \\
Unstable TOV   & 
  $1.4746$ & 
  $0.1843$ & 
  $4~(0)$ & 
  $(118.0, 59.0, 29.5, 14.7)$ & 
   $8.59$ & 
   $4.24$ & 
   $1.44$ &
  octant \\
BH*  & 
  $1.4746$ & 
  $0.1843$ & 
  $4~(0)$  & 
  $(118.0, 59.0, 29.5, 14.7)$ & 
   -- & 
  $3.41$ & 
  $1.44$ & 
  octant \\
BH  & 
  $1.4746$ & 
  $0.0921$ & 
  $5~(0)$ & 
  $(150.4, 82.6, 36.8, 18.4, 8.1)$ & 
  -- & 
  $2.36$  & 
  $1.00$ &
  octant \\
Binary NS (inspiral) & 
  $9.4510$ & 
  $0.2953$ & 
  $6~(2)$ & 
  $(756.1, 354.4, 177.2, 94.5, 32.5, 16.2)$ & 
  $12.4$   & 
  -- & 
  $1.57$  &
  $z$ and $\pi$  \\
Binary NS (merger)   & 
  $9.4510$ &
  $0.2953$ & 
  $6~(0)$ &
  $(756.1, 354.4, 177.2, 94.5, 37.8, 21.6)$ &
  -- &
  -- &
  -- & 
  $z$ and $\pi$  \\
Binary NS (collapse) & 
  $9.4510$ & 
  $0.1477$ & 
  $7~(0)$ &
  $(756.1, 354.4, 177.2, 94.5, 37.8, 21.6, 10.8)$ &
  -- & 
  $3.60$ &
  $3.20$  & 
  $z$ and $\pi$  \\
\end{tabular}
\end{ruledtabular}
\caption{Grid setups of our simulations. Here, $h_0$ and $h_f$ denote,
  respectively, the coarsest and finest grid spacing, $R_n$ are the radii
  of each refinement level (in case of levels with moving boxes, the
  radius of each box), $R_\text{star}$ is the initial neutron star
  coordinate radius, $R_\text{BH}$ the coordinate radius of the final BH,
  if present, and $M$ is the gravitational mass of the NS or of the
  BH. Note that the simulation BH* refers to a stationary BH evolved with
  a numerical domain with extents rescaled so that they match those of
  the collapsing star.}
\label{tab:grid}
\end{table*}

For our first comparison, we choose an isolated, spherically symmetric
NS. The initial data obey a polytropic EOS, \ie $P=K \rho^\Gamma$, with
$\Gamma =2$ and $K=100$. During the evolution, we use a matching ideal
gas EOS with $\Gamma =2$. The star has a (gravitational) mass of $M = 1.4
\usk M_\odot$ and a circumferential radius of $R= 14.16
\usk\kilo\meter$. The grid setup can be found in Table \ref{tab:grid}.

In order to add a well-defined initial constraint violation, we perturb
the star with an eigenfunction of the $\ell=2,m=0$ fundamental mode in the
Cowling approximation. The amplitude we used corresponds to a radial
velocity of $0.017$ at the surface. Because a corresponding perturbation
in the metric is not introduced, the Hamiltonian and momentum constraints
are violated in exactly the same way for CCZ4 and BSSNOK simulations.

When evolved in time, both formulations lead to stable solutions, and the
dynamics of the simulations agrees very well between BSSNOK and CCZ4,
with a relative difference in the central rest mass density that after
$7.1 \usk\milli\second$ is only $3\times 10^{-4}$. For comparison, the
amplitude of the oscillations corresponds to a relative change of
$0.015$. The constraint violations are shown in
\Fig{fig:tov_L2constr}. Since the components $M^i$ of the momentum
constraint are very similar, we show the combined norm $\Vert M \Vert_2
\equiv \sqrt{\sum_i \left(\Vert M^i \Vert_2 \right)^2}$. Clearly, the
Hamiltonian constraint is damped efficiently by the CCZ4 formulation
already after about 1 ms and is about 2 orders of magnitude smaller at
the end of the simulation, \ie at $t \simeq 7$ ms. The BSSNOK
formulation, on the other hand, exhibits a growth after a short initial
decrease. In this setup, the momentum constraints are on average very
similar for the CCZ4 and BSSNOK formulations.

\begin{figure}
  \centering
  \includegraphics[width=0.98\columnwidth]{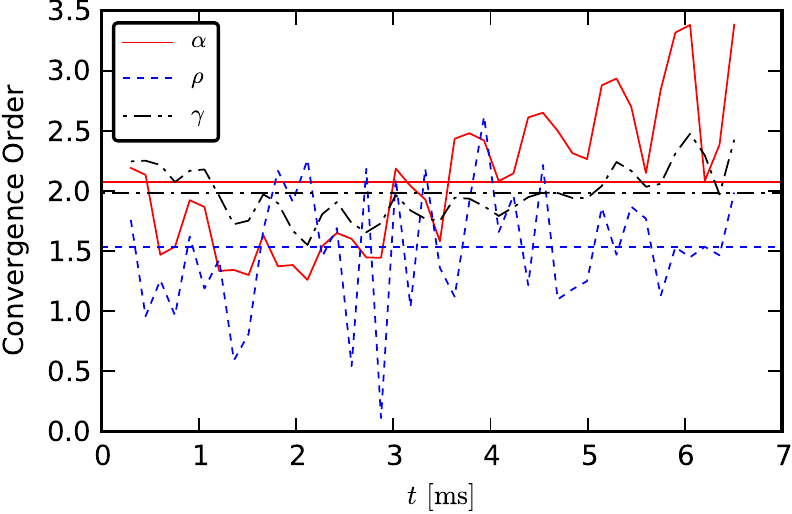}
  \caption{Convergence test of the implementation of the CCZ4 formulation
    when coupled to the matter evolution and when using a stable,
    perturbed TOV star. Shown is the time evolution of the convergence
    order obtained from the spatial $L_2$ norms of the residuals. The
    horizontal lines show the overall order obtained from the time-averaged residuals. }
  \label{fig:tov_conv_order}
\end{figure}

Besides investigating the constraints, we also use this setup to test
the convergence properties of the code. For this we use three different
grid spacings, \ie $\Delta x= 443.0, 295.3, 196.9 \usk\meter$ on the
finest level, differing by a factor of $1.5$, and evolved up to
$t=6.5\usk\milli\second$. For the variables $\alpha, \rho$, and
$\gamma$, we then measure the convergence order as follows:
First, we select the time steps that are common to all resolutions. Next,
we select the grid points on the finest refinement level which are present
for each resolution. Then we compute for each variable and time step the
differences between the results for low and medium resolutions, as well as
those between the results for medium and high resolutions. Finally, we compute the $L_1$, $L_2$, and
$L_\infty$ norms of those differences over the selected grid points, and
compute the time average of the norms. From those values we compute the
convergence order $n$, assuming the errors converge following a power
law. We also compute the convergence order obtained at each time, which
is shown in \Fig{fig:tov_conv_order}. For the lapse function, we find an
overall convergence order of $n=1.96,2.07,2.87$ for the $L_1$, $L_2$, and
$L_\infty$ norms, respectively, while for $\gamma$ we obtain
$n=1.83,1.98,2.62$. For the rest mass density $\rho$, on the other hand,
we find $n=1.45, 1.53, 1.85$.

The convergence order we would expect from the combined numerical schemes
for spacetime and fluid in regions where fluid quantities are smooth and
no fluid-vacuum boundary is present is 2. However, the convergence of
the hydrodynamic equations part is reduced near the surface of the star
because of the use of an artificial atmosphere; furthermore, the
convergence order also drops to 1 near shocks (although no global
shocks are present in this test) and at local maxima. Because of these
contaminations, the effective convergence order for the hydrodynamic
equations is smaller, around $n \simeq 1.5$. On the other hand, the
reason why the spacetime variables show better convergence in the given
resolution range is probably that the spacetime is mainly influenced by
the bulk properties of the matter, which are less affected by the
aforementioned problems. These results are in good agreement with what
was found when analyzing the convergence order of the waveforms from
binary NSs, which we will discuss in Sec.~\ref{binaries} (see also Ref.~\cite{Baiotti:2009gk}).

\subsection{Collapse to a black hole}
\label{sec:collapse}

Our next test consists of a NS on the unstable branch, to which we apply
a small inwards velocity kick in order to trigger a collapse to a black
hole. We use the same EOS as for the stable TOV test in the previous
section, but the central density is $5 \times 10^{15}\,{\rm g~cm}^{-3}$,
the gravitational mass is $M = 1.44 \Msun$, and the circumferential
radius is $R=8.59\usk\kilo\meter$. The velocity perturbation is simply
given by $v^r = \mathcal{V} r/R$, with $\mathcal{V} = -0.01$. The purpose
of this test is not only to compare the constraint violations between the
CCZ4 and the BSSNOK formulations, but also to show that both formulations
lead to a stable evolution of the black hole when coupled to matter.

We observe that the infalling matter ends up in the central numerical
cell. The matter then stays there; the relative change of total baryon
mass between $t=1.9 \usk\milli\second$ (after the collapse) and $5
\usk\milli\second$ is less than $10^{-7}$. The profile of the lapse
function, shift vector, density, and metric determinant all approach
stationary values shortly after the BH has formed. Between $1.9$ and $5
\usk\milli\second$, the lapse function at the center changes by less than
$3 \%$. There is, however, a numerical instability in the fluid velocity
inside the cell at the center of the BH, but since we limit the maximum
velocity at the center of the BH as described in Sec.~\ref{sec:num_meth}, this quickly becomes stationary as well.

\begin{figure}
  \centering
   \includegraphics[width=0.98\columnwidth]{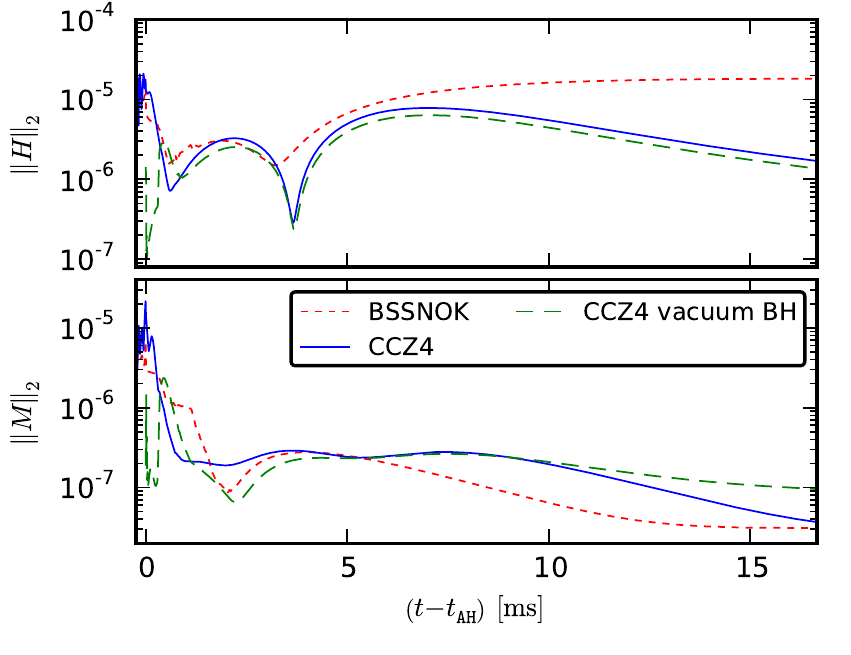}
   \caption{Comparison of the CCZ4 and BSSNOK simulations for the
     collapse of an unstable TOV star to a BH. Shown are the $L_2$ norms
     of the Hamiltonian (top panel) and the momentum constraint violations
     (bottom panel), with the norms taken over the region outside the apparent
     horizon. The constraint violations are given in units
     $G=c=M_\odot=1$. For comparison, we also show results for a pure
     vacuum CCZ4 simulation of a single Schwarzschild BH of the same mass
     as the NS model and using the same grid setup. The parameters for
     the CCZ4 simulations are $\kappa_1=0.065,\, \kappa_2=0,\,
     \kappa_3=0.5$. The time coordinate is relative to the time
     $t_\mathtt{AH}$ of apparent horizon formation. }
  \label{fig:collapse_constr}
\end{figure}

The evolution of the constraints is shown in \Fig{fig:collapse_constr}.
For comparison, we evolved a Schwarzschild BH of the same mass using the
CCZ4 spacetime evolution code without the fluid part, but with the same
grid setup. As shown in \Fig{fig:collapse_constr}, the constraint
violations are very similar to the collapse simulation after the BH has
formed. The $L_2$ norms excluding the BH interior agree very well, not
only in magnitude but also in terms of the long-time behavior.

\begin{table}
\begin{ruledtabular}
\begin{tabular}{llccccc}
Type&
Formulation&
$\kappa_1$&
$\kappa_3$&
$h_f / M$&
$L / M$ &
$\delta M/M$
\\\hline 
BH&
CCZ4&
$0.020$&
$0.5$&
$0.0625$&
$102.0$&
$0.16$ $\%$
\\
BH&
CCZ4*&
$0.020$&
$1.0$&
$0.0625$&
$102.0$&
$2.80$ $\%$
\\
BH&
CCZ4*&
$0.100$&
$1.0$&
$0.0625$&
$102.0$&
$0.10$ $\%$
\\
BH&
CCZ4&
$0.065$ & 
$0.5$&
$0.0866$&
$55.4$&
$0.76$ $\%$
\\
BH&
Z4c&
$0.020$&
--&
$0.0625$&
$102.0$&
$0.03$ $\%$
\\
Collapse&
BSSNOK&
--&
--&
$0.0866$&
$55.4$&
$0.03$ $\%$
\\
Collapse&
CCZ4*&
$0.100$&
$1.0$&
$0.0866$&
$55.4$&
$1.01$ $\%$
\\
Collapse&
CCZ4*&
$0.170$&
$1.0$&
$0.0866$&
$55.4$&
$0.23$ $\%$
\\
Collapse&
CCZ4&
$0.020$&
$0.5$&
$0.0866$&
$55.4$&
$2.09$ $\%$
\\
Collapse&
CCZ4&
$0.065$&
$0.5$&
$0.0866$&
$55.4$&
$0.64$ $\%$
\\
Collapse&
CCZ4&
$0.100$&
$0.5$&
$0.0866$&
$55.4$&
$2.21$ $\%$
\\
Collapse&
CCZ4*&
$0.100$&
$0.5$&
$0.0866$&
$55.4$&
$0.51$ $\%$
\\
Collapse&
Z4c&
$0.020$&
--&
$0.0866$&
$55.4$&
$0.16$ $\%$
\end{tabular}
\end{ruledtabular}
\caption{Accuracy of the BH mass for simulations of nonrotating vacuum
  BHs and BHs created by the collapse of unstable spherical NSs. Note
  that the simulations marked with $*$ use the prescription
  Eq.~\eqref{eq:new_kappa_1} for the damping coefficient $\kappa_1$, in
  particular those performed with the covariant formulation, \ie
  with $\kappa_3=1$. The maximum deviation of the measured BH mass from
  the exact value during the first $800\usk M$ after the detection of the
  apparent horizon is denoted by $\delta M$. For the collapse
  simulations, the time up to $200\usk M$ after BH formation is
  ignored. $L$ is the position of the outer boundary, and $h_f$ the grid
  spacing on the finest level.}
\label{tab:bh_mass}
\end{table}

The norms including the BH interior (not shown in the plot) for vacuum BH
and collapse agree very well initially. However, when the aforementioned 
instability develops during the evolution with matter, the norm of the 
momentum constraint is increased by a factor $\approx 2$. In order to 
determine whether this affects the exterior of the BH,
we conduct a numerical experiment. Directly after the formation of the BH, 
we introduce a source term to the fluid variables as described in 
Sec.~\ref{sec:numsetup} in order to exponentially remove the matter over
an e-folding timescale $\tau_d=0.025 \usk\milli\second$.
This source term is only active near the center of the BH, 
which we define for simplicity via the lapse function by the condition 
$\alpha<0.007$. The instability at the center of the BH and the corresponding
jump in the momentum constraint do not occur anymore when using this method. 
However, there is no visible change of the norm of the constraint violations 
outside the BH. 

We conclude that the constraint violations outside the BH are not
influenced significantly by the presence of matter inside the BH. The
coupling of the hydrodynamic evolution to the spacetime evolution with
the CCZ4 method does not seem to compromise the stability of the BH
evolution in any way. It is, however, necessary either to limit the fluid
state at the center of the BH as described in Sec.~\ref{sec:num_meth}
or to gradually remove the matter from the center as described above.

As a measure of accuracy, we monitor the BH mass extracted using the
isolated horizon formalism. For the collapse of a spherical star, the BH
has to be stationary after all matter has crossed the horizon, and the BH
mass has to be exactly the ADM mass of the initial unstable star. We
therefore compute the maximum deviation of the numerically extracted
value from the exact one during the time interval $200$--$800 \usk M$
after apparent horizon detection. The results are reported in Table
\ref{tab:bh_mass}. For the noncovariant CCZ4 simulations of the
collapse, the accuracy is around $0.64$--$2.2\usk\%$ for damping
parameters $0.02 \le \kappa_1 \le 0.1$. As we will show in
Sec.~\ref{sec:Cvsc}, the error can be reduced below $0.3\%$ by a
modification of the damping terms, which allows the use of the covariant
version ($\kappa_3=1$). Nevertheless, the standard BSSNOK and Z4c
formulations are more accurate in terms of BH mass for this test.

\subsection{Binary neutron stars with constraint-violating initial data}
\label{sec:merger}

\begin{figure}
  \centering
  \includegraphics[width=0.98\columnwidth]{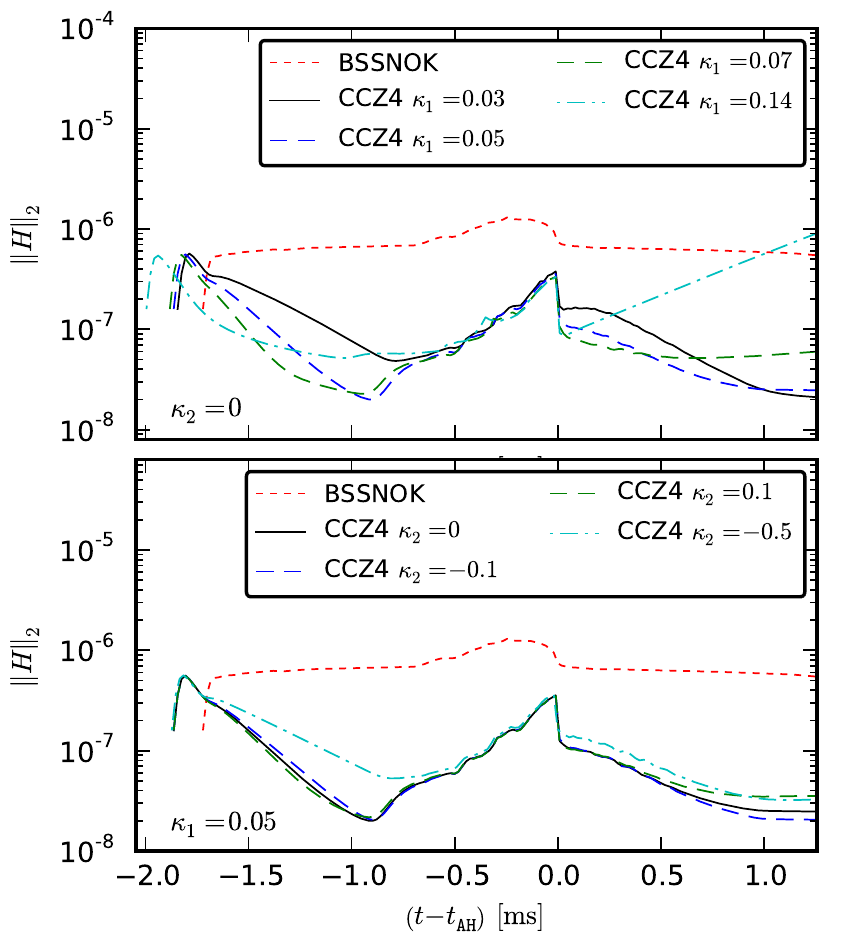}
  \caption{Comparison of the Hamiltonian constraint violations when
    evolving the eccentric binary, using the CCZ4 system with different
    parameters, and with the BSSNOK system. \textit{Top panel:} Influence
    of parameter $\kappa_1$, keeping $\kappa_2=0$, $\kappa_3=0.5$ fixed.
    \textit{Bottom panel:} Influence of parameter $\kappa_2$, keeping
    $\kappa_1=0.05$, $\kappa_3=0.5$ fixed. Shown is the time evolution of
    the $L_2$ norm of the Hamiltonian constraint, excluding the interior
    of the apparent horizon. }
  \label{fig:eccbns_L2H}
\end{figure}

We next present tests of the CCZ4 formulation when applied to merging
binary NSs with constraint-violating initial data. These are
probably the most demanding tests with respect to constraint violations
because the initial data already contain large violations of the
constraint equations.

We start by considering binaries on eccentric orbits and set up such
initial data by starting with a constraint-satisfying solution
representing an irrotational binary on a quasicircular orbit and then
simply scale the velocity by a factor of $0.85$ to make the orbit eccentric
(which also speeds up the inspiral). This naturally introduces a large
constraint violation, thus allowing us to compare the evolution schemes
under extreme but well-defined conditions. More specifically, the
original initial data represent an irrotational binary system on a
quasicircular orbit, with equal baryon masses of $M_b=1.779\usk
M_\odot$, and an initial separation of $d=45\usk\kilo\meter$. The stars
obey a polytropic EOS with $\Gamma=2$ and $K=123.629$, while during the
evolution we use an ideal gas EOS with $\Gamma=2$. This model is publicly
available from the \texttt{LORENE} code.

We evolve the eccentric system with the CCZ4 formulation, using various
combinations of the parameters, as well as with the BSSNOK formulation.
The rest of the setup stays exactly the same, in particular the gauge
conditions. Figure \ref{fig:eccbns_L2H} shows the evolution of the
Hamiltonian constraint. As one can see, the constraint violations, which
are the same initially, are reduced by up to 1 order of magnitude
during the inspiral when using the CCZ4 formulation. The BSSNOK
formulation, on the other hand, shows a moderate growth during the
inspiral. Of course, the impact of the initial constraint violation on
the actual dynamics of the binary is difficult to assess. Any constraint
violation is obviously a deviation from the solution of the Einstein
equations, although the quantitative relation between the constraint
violations and the error on the physical quantities is largely
unknown. It is, however, reasonable to assume that a reduction of
constraint violations will also lead to more accurate results for the
physical quantities. In this respect, the CCZ4 formulation is clearly
better for the case considered here. Note that the accuracy that can be
achieved in this way is still limited by the constraint-satisfying
component of the evolution error, so that 
a further reduction will not increase significantly the overall accuracy.

In order to assess how the constraint violation influences the orbital
motion, we have compared the trajectories of the two NSs obtained with the
CCZ4 and BSSNOK formulations by tracking the ``barycenters'' of the two
NSs. \footnote{We define the position of the ``barycenter'' of each NS 
  simply as an extension of the Newtonian expression, \ie
  $\left(\int_V \hat{D} x^i \,\mathrm{d}^3x \right)
   \left( \int_V \hat{D} \,\mathrm{d}^3x \right)^{-1}$, where the 
   integration volume $V$ is suitably chosen to fully contain the 
   selected NS, but exclude the other.} We
find that significant deviations between the two trajectories develop
during the evolution while simultaneously the differences in the
constraint violations grow. After one orbit, the coordinate separation
already differs by $18\usk\%$. This is to be expected. Once the
constraints are violated, the numerical solution of our evolution system
belongs to an extended set of solutions of the Einstein equations. Even
starting with the same amount of constraint violation, the evolution
equations for the CCZ4 and BSSNOK formulations are expected to lead to
slightly different results, in the vicinity of the true Einstein
solution. The only question is below which magnitude of the constraint
violation the errors become tolerable. Clearly, the constraint violation
introduced by the crude rescaling of the velocity is already too large to
obtain meaningful results.

\begin{figure}
  \centering
  \includegraphics[width=0.98\columnwidth]{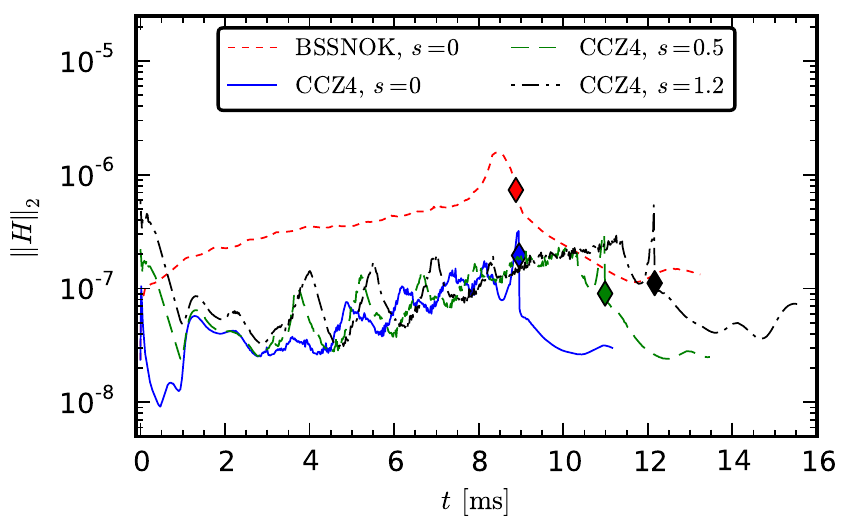}
  \caption{Evolution of the Hamiltonian constraint during mergers of
    spinning binary NSs on quasicircular orbits, where the spin is
    added to the initial data in a constraint-violating way. Shown are
    results obtained with the CCZ4 formulation for various amounts of
    spin. For the irrotational model, the results of the BSSNOK
    formulation are also shown; the symbols mark the time of formation of
    the apparent horizon.}
  \label{fig:bns_spin}
\end{figure}

When comparing simulations with CCZ4 using different parameters, we find
that increasing the damping parameter $\kappa_1$ leads to a decrease in the
violations of the Hamiltonian constraint during the inspiral phase. This
is indeed what one would naively expect, given that a larger $\kappa_1$
amounts to a smaller timescale for the damping of the constraint
violations. However, for $\kappa_1 \gtrsim 0.07$, we find an exponential
growth after the BH has formed. Hence, the optimal choice for a stable
evolution seems to be $\kappa_1 = 0.05$. Note that this value should scale
with the inverse mass of the BH; see Sec.~\ref{sec:CCZ4}. So far, we have
used a damping parameter $\kappa_2=0$. We did not find any significant
improvement by trying different values. On the contrary, for $\kappa_2 =
-0.5$ the constraint damping becomes less efficient.

Next, we consider mergers of rotating NSs. Because self-consistent
initial data for rotating NS binaries are not available (although an
approach to computing such models has been proposed recently in
Ref.~\cite{Tichy12}), we will show in the following that is possible
to use a short evolution with the CCZ4 scheme to convert
constraint-violating initial data into self-consistent initial data. This
allows us to study the influence of the additional NS spins on the spin
of the final BH and the surrounding disk. Although some of these results
have already been presented in Ref.~\cite{Kastaun2013}, for completeness,
we review here the behavior of the two formulations in this illustrative
test.

The models we investigate are constructed in a similar way as for the
eccentric case, namely, by starting from irrotational quasicircular
initial data. The spin is added by simply rescaling the velocity field
in the co-orbiting frame by a factor $1-s$, where $s=0$ corresponds to
the original irrotational model, and $s=1$ roughly to the corotating one
(see Ref.~\cite{Kastaun2013} for details). Naturally, this introduces
constraint violations, which are, however, not as strong as for the
eccentric binary. It also causes ordinary oscillations of the star, which
affect only the realism of the initial conditions and are not important
within the scope of this test.

The spatial distribution of the constraint violation shown in Ref.~\cite{Kastaun2013} is similar to that shown in \Fig{fig:qcbns_Hxy},
only the initial constraint violation is more pronounced. Figure
\ref{fig:bns_spin} shows the evolution of the Hamiltonian constraint for
various amounts of spin, with the symbols marking the time of
formation of the apparent horizon. The increase of the initial constraint
violation with the spin is clearly visible. However, even for the
fastest-spinning model, the evolution with the CCZ4 formulation rapidly
reduces the Hamiltonian constraint to a much smaller magnitude, which
already after ${\sim}~1\usk\milli\second$ becomes comparable with the one
obtained with the BSSNOK formulation when evolving constraint-satisfying
initial data. As we will show in Sec.~\ref{binaries}, the amount of
constraint violations for the latter are tolerable. We thus conclude that
at these separations an evolution time of ${\sim}~1 \usk\milli\second$
with the CCZ4 formulation is sufficient to produce initial data that can
be considered self-consistent. Of course, although self-consistent, the
initial data may well represent a physical system which is rather
different from the intended one.

\begin{figure}
  \centering
  \includegraphics[width=0.98\columnwidth]{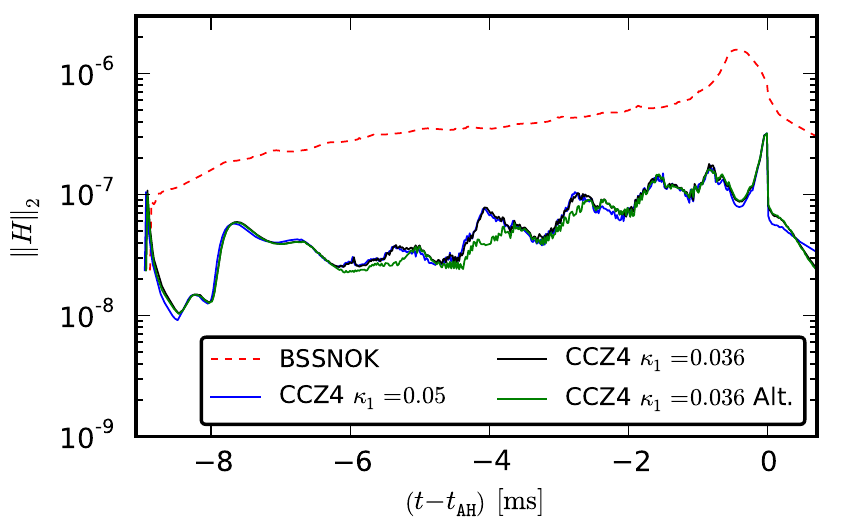}
  \caption{Comparison of the Hamiltonian constraint violations when
    evolving with the BSSNOK and CCZ4 systems, for the quasicircular
    coalescence. Shown is the time evolution of the $L_2$ norm of the
    Hamiltonian, excluding the interior of the apparent horizon. For CCZ4,
    we show results obtained with $\kappa_1=0.05$, $\kappa_1=0.036$, and
    $\kappa_1=0.0036$ using an alternative grid structure (``Alt.'')
    which contains moving boxes only on the finest level. All CCZ4 runs
    were carried out with $\kappa_2=0, \kappa_3=0.5$.}  \label{fig:bns_L2H}
\end{figure}

\begin{figure*}
  \centering
  \includegraphics[width=2.0\columnwidth]{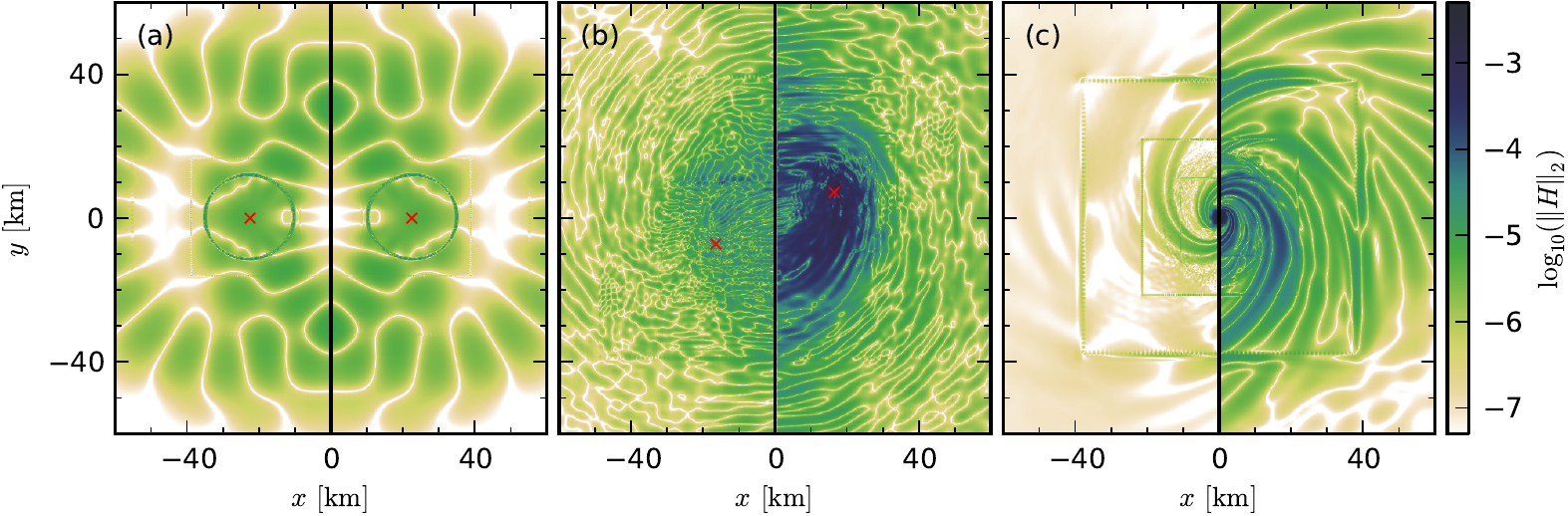}
  \caption{Comparison of the local Hamiltonian constraint violations when
    evolving the quasicircular coalescence with the BSSNOK and CCZ4
    systems. The panels depict the Hamiltonian constraint in the $(x,y)$
    plane at different times: (a) $t=0$, initial data, (b)
    $t=6\usk\milli\second$, inspiral, (c) $t=12\usk\milli\second$, final
    state. The left half of each panel shows the CCZ4 results, and the right
    half shows the BSSNOK results. The locations of the NS barycenters are
    marked by the red crosses.}
  \label{fig:qcbns_Hxy}
\end{figure*}

\subsection{Binary neutron stars with constraint-satisfying initial data}
\label{binaries}

We now turn to investigate the behavior of merging binary NSs in
quasicircular orbits. For this, we evolve the original binary star model
described in the previous subsection, without reducing the linear
momenta. Again, we find that the CCZ4 formulation suppresses the
Hamiltonian constraint by roughly 1 order of magnitude when compared to
the BSSNOK formalism, as shown in \Fig{fig:bns_L2H}. Varying the damping
constant $\kappa_1$ in the stable range impacts the results only
marginally. Figure \ref{fig:bns_L2H} also shows a clear periodic
increase/decrease of the Hamiltonian constraint violation when using the
CCZ4 formulation. We believe that this behavior is related to the
movement of the refined boxes. Whenever a refined box moves, the new
points have to be computed by interpolation from the coarser grid, which
introduces an additional error. Indeed, the period of the variations in
the constraint violations corresponds to half an orbit, which is
compatible with the $\pi$ symmetry of the binary. However, to further
validate this hypothesis, we perform a simulation where only the finest
level consists of moving boxes, in contrast to the standard setup where
the two finest levels are moving boxes. As one can see in
\Fig{fig:bns_L2H} by comparing the solid black and green lines, this has
some influence during inspiral. Moreover, it is most prominent at the
stage where the change of the overlap of the moving boxes on the second-finest level is also large. The fact that we do not observe the periodic
pattern for the BSSNOK results seems to indicate that the error due to
regridding is not dominant in this case.

Since binary NS mergers are an important application of our code, we
perform a computationally expensive convergence test with three
resolutions, each increased by a factor of $1.5$. We measure the
convergence order by the same method used for the single star, only that
we use the finest nonmoving refinement level instead of the finest
one. Furthermore, we exclude the interior of the apparent horizon (more
precisely, we exclude a coordinate sphere with the mean coordinate radius
of the horizon). For the lapse function, we measure an overall convergence
order of $n=1.9$, while for the rest mass density this is $n=1.6$, all
measured using the $L_2$ norm. The metric determinant $\gamma$ develops a strong
peak at the center of the grid shortly before the apparent horizon
forms. This is normal, but complicates measuring the convergence after BH
formation. Indeed, during the inspiral, we obtain a convergence order
$n=1.7$ for $\gamma$, which is recovered again after the BH formation. 
We note that the time-dependent estimate for the convergence order of all variables fluctuates
strongly during the merger (up to $n=8$), which is probably caused by the
accumulated phase error. As a consequence, we cannot prove convergence
for this stage. We can, however, establish convergence of the end result,
\ie of the final BH, which should be insensitive to the phase error. To
this end, we compare the final mass and spin of the BH as well as the
time until apparent horizon formation.

\begin{figure*}
\centering
  \includegraphics[width=2.0\columnwidth]{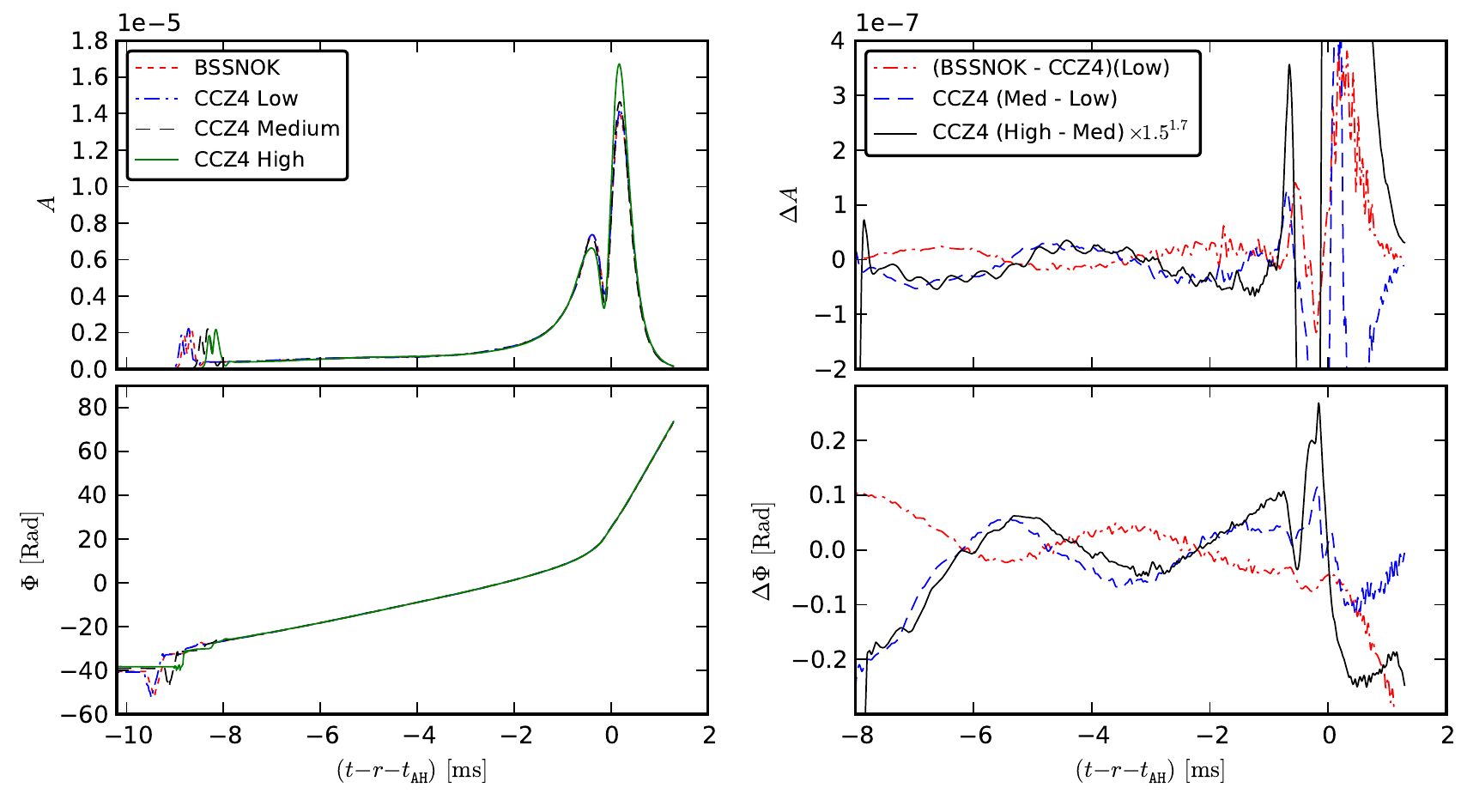} \caption{Accuracy
  of the gravitational wave signal in terms of the $\ell=m=2$
  multipole component of the Weyl scalar $\Psi_4$, extracted at
  $r=664\usk\kilo\meter$. \emph{Top left:} Amplitude of $\Psi_4$.
\emph{Bottom left:} Complex phase $\Phi$ (see main text). \emph{Top
right:} Residuals of the amplitude between CCZ4 simulations with
different resolutions, and between BSSNOK and CCZ4 at lowest resolution.
The residual between high and medium resolution is rescaled assuming a
convergence order of $1.7$. \emph{Bottom right:} Residuals of the
phase, also rescaled.}  \label{fig:bns_gw}
\end{figure*}

As reported in Table \ref{tab:bns_phys_quant}, we find a convergence
order around $1.2$ for the BH properties. This is not surprising since
the convergence order of the hydrodynamic evolution scheme probably
reduces to 1 during the merger due to the formation of strong
shocks. Nevertheless, the errors of BH mass and spin are quite
small. Using Richardson extrapolation, we obtain an extrapolated
total error for the values obtained at the lowest resolution, which is
$0.9\usk\%$ for the BH mass and $0.3\usk\%$ for the spin. The appearance
of an apparent horizon is the result of an independent search algorithm
on a complex combination of the evolved quantities and, as such, not
necessarily sharing the same convergence order of the evolved
equations. That said, we find that the time of first appearance of an
apparent horizon has a convergence order $n=2.1$, but is also the
quantity with the largest error. The error with respect to the Richardson
extrapolated time is $9\usk\%$ for the lowest resolution and $2 \usk \%$
for the highest. The error introduced by the different frequency at which
the apparent horizon is searched at different resolutions is much smaller
and only $\sim~ 0.1\usk\%$. 

\begin{table}
\centering
\begin{ruledtabular}
\begin{tabular}{lllccc}
& BSSNOK      
& CCZ4      
& $\Delta_\text{Form}$                          
& $\Delta_\text{Res}$       
& $n$
\\\hline
$M_{_\mathtt{BH}} / M_\odot$ & 
$3.222$ & 
$3.204$ & 
$0.6 \usk\%$ & 
$0.5\usk\%$ & 
$1.2$ 
\\
$a_{_\mathtt{BH}}$ & 
$0.837$ & 
$0.840$ & 
$0.4 \usk\%$  & 
$0.2\usk\%$ & 
$1.2$ 
\\
$t_{_\mathtt{BH}}~[\milli\second]$ & 
$8.88$ & 
$8.94$ & 
$0.7 \usk\%$ & 
$7.0\usk\%$ & 
$2.1$ 
\\
\end{tabular}
\end{ruledtabular}
\caption{Comparison of physical quantities at the end of the
  quasicircular coalescence,obtained with the BSSNOK and CCZ4
  formulations. Above, $M_{_\mathtt{BH}}$ is the mass of the final BH,
  $a_{_\mathtt{BH}}=J_{_\mathtt{BH}}/M_{_\mathtt{BH}}^2$ its
  dimensionless spin parameter, and $t_{_\mathtt{BH}}$ is the time when
  the apparent horizon is detected. The difference between the highest
  and lowest resolution of the CCZ4 convergence test is given by
  $\Delta_\text{Res}$, while $p$ is the measured convergence order. The
  difference between results obtained with the BSSNOK and CCZ4
  formulations at the lowest resolution is denoted by
  $\Delta_\text{Form}$. }
\label{tab:bns_phys_quant}
\end{table}

In contrast to the eccentric case, the dynamics of the quasicircular
system agrees well between the CCZ4 and BSSNOK formulations. To quantify
this statement, we compare the mass, spin, and formation time of the BH in
Table~\ref{tab:bns_phys_quant}. The agreement is better than
$1\usk\%$. The differences in the BH properties are thus comparable to
the numerical errors of the CCZ4 simulation as determined by the
convergence test. Curiously, the difference in horizon formation time is
an order of magnitude smaller than the numerical errors for this
particular case, although we do not expect that this holds in
general. Since the constraint violations still differ by a factor
$\approx 10$ between CCZ4 and BSSNOK, one can conclude that the magnitude
of constraint violations observed in the quasicircular BSSNOK evolution
is tolerable, while the amount present in the eccentric case already
leads to severe errors. This is a very useful notion since, as we discussed
before, the relation between the constraint violation and error of physical
quantities is largely unknown. We stress that the $L_2$ norm we use to
quantify the constraint violations depends on the computational volume
and the falloff behavior of the constraint violations. For different
setups, one has to rescale accordingly in order to make sensible
comparisons.

Besides the amount of constraint violation, we are also interested in its
spatial distribution. Figure \ref{fig:qcbns_Hxy} shows the Hamiltonian on
a cut in the $(x,y)$ plane. Clearly, most of the violations are produced
along the orbit of the stars. Also, the mesh refinement boundaries are
clearly visible. In practice, the stars leave behind a trail of
constraint violations, which is more pronounced and decays more slowly
for the BSSNOK formulation. Furthermore, constraint violations travel
through the computational domain and are partially reflected at each
refinement boundary, for both the CCZ4 and the BSSNOK formulations. After
the formation of the BH, when the system approaches a final state, the
constraints exhibit a relatively stationary spiral pattern of slowly
decreasing magnitude.

Finally, we should also remark that the relative difference in the
constraint violations that we measure in the simulations reported here
refers to \emph{our} implementation of the CCZ4 and BSSNOK formulations
and should not be considered as universal. Because a number of slightly
different versions of the BSSNOK system are used by different groups, it
is possible that the differences between the two formulations could also
be smaller or larger when performed by other groups.

\subsubsection{Gravitational wave signal}

One of the most important results of binary NS merger simulations is
obviously the emission of gravitational wave (GW) signals. Since prior to extraction,
GWs travel into the weak-field region, crossing several refinement
boundaries and becoming a small perturbation due to the $1/r$ falloff, they
might be affected more strongly by numerical errors than the bulk
dynamics of the merger. In particular, constraint violations could affect
the GWs differently. In the following subsection, we measure the numerical accuracy
of the GW signal using the CCZ4 convergence test and estimate the
influence of constraint violations by comparing between the CCZ4 and the
BSSNOK formulations.

For this, we extract the $\ell=m=2$ component of the Weyl scalar $\Psi_4$
at a fixed radius $r=664\usk\kilo\meter$. We then decompose the complex
quantity $\Psi_4$ into amplitude and phase, \ie $\Psi_4 = A \exp(i
\phi)$, where $\phi$ is a continuous function of time. To compare
different simulations, we measure time with respect to the time
$t_{_\mathtt{AH}}$ at which the apparent horizon forms, and introduce the
phase difference $\Phi(t)= \phi(t) - \phi(t_{_\mathtt{AH}})$. The
amplitude and the phase are shown in the left panels of \Fig{fig:bns_gw}
for the BSSNOK and CCZ4 results. For the CCZ4 simulations, we plot the
three resolutions of the convergence test, where the lowest one is
identical to the one used for the BSSNOK simulation. One can clearly
distinguish the inspiral phase, a first peak corresponding to the merger,
and a second peak corresponding to the ringdown of the BH. The usual junk
radiation inherent to the initial data can also be seen at the beginning
of the evolution. In the right panels of \Fig{fig:bns_gw}, we show the
residuals of amplitude and phase between different resolutions and
between the CCZ4 and BSSNOK runs when performed at the lowest
resolution. For the CCZ4 runs, we find errors compatible with a
convergence order around $1.7$ during the inspiral. During the merger, on
the other hand, the convergence decreases and is lost during the ringdown
phase. From the time derivative of $\Phi$, we compute the instantaneous
frequency, which increases from around $\dot{\Phi} \approx 7
\usk\mathrm{rad}~\milli\second^{-1}$ at the end of the inspiral to
$\dot{\Phi} \approx 35 \usk\mathrm{rad}~\milli\second^{-1}$ at the
maximum of the ringdown signal. Unfortunately, the wavelength
corresponding to the latter is resolved by only six coarsest grid points
of the lowest resolution, and thus the signal from the ringdown is
severely under-resolved. In order to demonstrate convergence of the
high-frequency part of the signal, we would have to repeat the runs with
much higher resolution in the weak-field region.

\begin{figure}
  \centering
  \includegraphics[width=0.98\columnwidth]{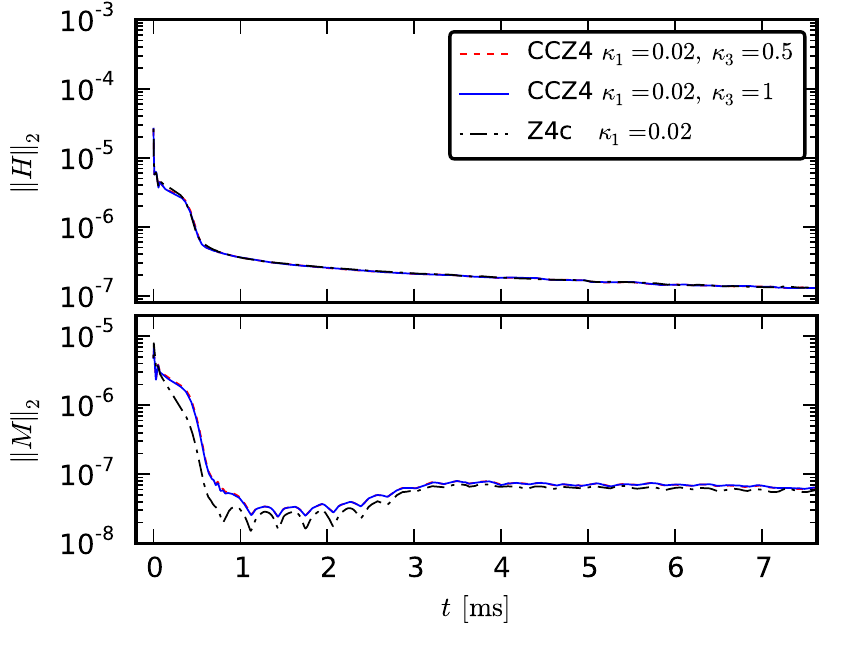}
  \caption{Comparison between the noncovariant and fully 
    covariant CCZ4 and the noncovariant Z4c systems in evolutions of a 
    nonrotating stable neutron star. Shown is the time evolution of the 
    $L_2$ norm of the Hamiltonian constraint violation (top panel) and 
    the momentum constraint violation (bottom panel).}
  \label{fig:tov_constr}
\end{figure}

In the analysis of GW data using matched filtering techniques, the most
important error is the phase shift during the inspiral. For our test
case, we find that the total phase error until the merger is less than $0.4
\usk\mathrm{rad}$ for the lowest resolution, and $0.1 \usk\mathrm{rad}$
for the highest one. At first sight, this seems to contradict the larger
relative error we obtain for the time until BH formation (compare
Table~\ref{tab:bns_phys_quant}). However, by analyzing the coordinate
separation\footnote{Note that although the gauge condition is the same
  for both simulations, the coordinate separation is a gauge-dependent
  quantity.} of the stars barycenter, we find that the relative error of
separation is larger than the error of the GW phase; \ie the orbital
period is more accurate than the decrease in separation per orbit. This
is reflected in the relatively large amplitude error (see
\Fig{fig:bns_gw}), since the GWs at the same phase (or time) are produced
at different orbital separations.

Figure \ref{fig:bns_gw} also shows that the differences in the waveforms
obtained with the CCZ4 and the BSSNOK formulations are generally
comparable to the numerical errors, even during merger and
ringdown. Since the constraint violations differ by 1 order of
magnitude, we can conclude that their impact on the GW signal is also
comparable to that due to ordinary numerical errors, or even smaller
(the differences could as well be purely due to numerical errors other
than the constraint violation). This result is rather reassuring since,
to the best of our knowledge, the impact of the constraint violations on
the accuracy of the GW signal has not been measured before.

\subsection{Comparison with the Z4c formulation}
\label{sec:z4c}

\begin{figure}
  \centering
  \includegraphics[width=0.98\columnwidth]{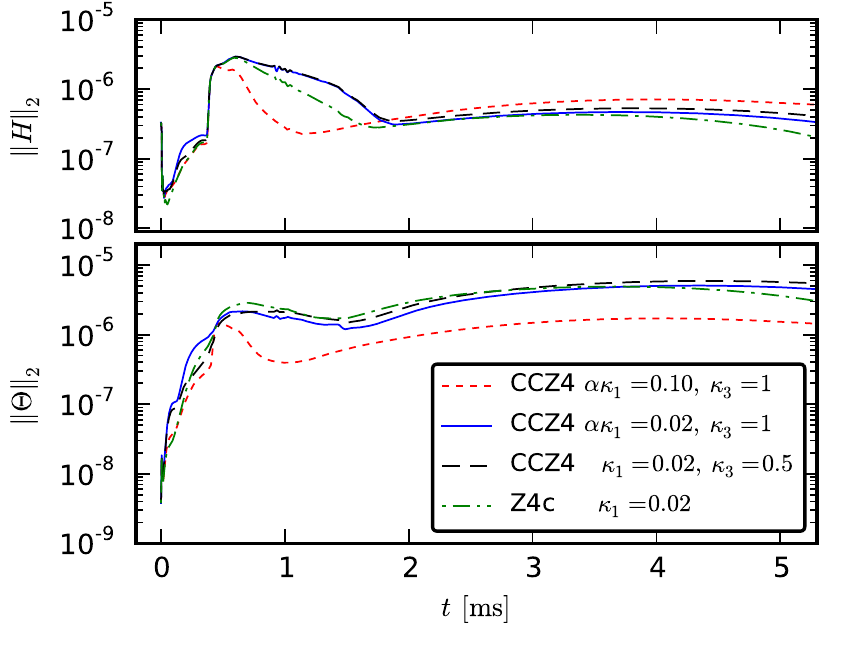}
  \caption{Comparison between the noncovariant and covariant versions 
    of the CCZ4 and the noncovariant Z4c systems in evolutions of a
    nonrotating BH. Shown is the time evolution of the $L_2$ norm of the
    Hamiltonian constraint violation (top panel) and of the $\Theta$ variable
    (bottom panel), computed excluding the interior of the BH. }
  \label{fig:sbh_constr}
\end{figure}

As anticipated in the Introduction, we now present a comparison of the
results obtained with another conformal formulation of the Z4 system,
namely the Z4c formulation proposed in Ref.~\cite{Hilditch2012}. To
reduce the computational costs, the comparison will be carried out with
the evolutions of a stable nonrotating star, of a single nonrotating BH
and of an unstable NS which collapses to a BH. However, we expect that
the qualitative behavior of the two formulations will extend also to
binary systems either of BHs or of NSs. As far as the Z4c system is
concerned, we have implemented the formulation described in Ref.~\cite{Hilditch2012} within the \texttt{McLachlan}
code~\cite{Brown2007b}. This requires only minor modifications in the
source terms of the CCZ4 system (see Sec.~\ref{sec:CCZ4} for details
about the differences between the two systems). As mentioned in
Sec.~\ref{sec:numsetup}, we use the radiative boundary conditions
provided by the \texttt{McLachlan} code.

We note that both the CCZ4 and the Z4c formulations implement the same
constraint-damping scheme~\cite{Gundlach2005:constraint-damping}, and we
use the values of the constraint-damping parameters advocated as best
suited for the Z4c formulation in Ref.~\cite{Hilditch2012}, namely
$\kappa_1=0.02$ and $\kappa_2=0.0$. Also, for both systems we monitor the
same quantities, namely the behavior of the Hamiltonian and momentum
constraint violations, but also of the $\Theta$ function, whose time
variation measures the size of the Hamiltonian constraint violation and
thus assesses the deviation of the numerical solution from the true
Einstein solution. Indeed, we find this diagnostic quantity to be a very
important indicator of the quality of the solution, which can be compared
directly for the two conformal formulations (we recall that the $\Theta$
function is not defined for the BSSNOK formulation).

The first test involves a nonrotating stable NS, as described in Sec.~\ref{sec:isolated}. Overall, in the presence of matter the two systems
provide an almost identical behavior and show a 2-order-of-magnitude
decrease in the constraint violations as the evolution is started. These
are reported in Fig. \ref{fig:tov_constr}, which shows both the
violations in the Hamiltonian and the momentum constraints (top and
bottom panels, respectively). Note that the Z4c evolution has a slightly
larger violation of the Hamiltonian constraint (top panel) and a smaller
one in the momentum constraints (bottom panel). Overall we find the two
violations comparable.

The second test involves the evolution in vacuum of a
nonrotating BH. The results are again very similar for the two systems,
as one can see in Fig. \ref{fig:sbh_constr}, which reports the violation
in the Hamiltonian constraint and the evolution of the $\Theta$ function
(top and bottom panels, respectively). Note that while the Hamiltonian
constraint is slightly smaller for the Z4c formulation, the momentum
constraint is smaller for CCZ4. Both systems show small deviations in the
final BH mass; see Table~\ref{tab:bh_mass}.

The third comparative test consists in the collapse of an unstable TOV
star, as described in Sec.~\ref{sec:collapse}. Figure~\ref{fig:tovcoll_constr} shows again the violation in the Hamiltonian
constraint and the evolution of the $\Theta$ function (top and bottom
panels, respectively). The only significant difference in this case is a
spike in the CCZ4 formulation right after collapse, which is efficiently
damped after $1 \usk\milli\second$ of evolution, so that the constraint
violations return to values similar to the Z4c formulation. The value of
the BH mass after collapse shows larger errors for CCZ4, around $2.09\%$,
in comparison with only $0.16\%$ for Z4c (see Table
\ref{tab:bh_mass}). However, this result can be improved to $0.51\%$ error
by using a modified damping scheme with $\kappa_1 = 0.1$ (see Sec.~\ref{sec:Cvsc}).

Based on the results presented above, we conclude that the CCZ4 and the
Z4c formulations yield very similar results in terms of their ability to
damp the violations in the constraint equations. However, one important
difference remains between the two systems that is, only the CCZ4
formulation with $\kappa_3=1$ represents a version of the original
Z4 system that is not only conformal but also \emph{covariant}. As a
result, only for this covariant formulation should one reasonably expect
that the qualitative behavior of the constraint violations will remain
unchanged when evolving the same system in different coordinates.
Finally, given the similar behavior of the CCZ4 and of the Z4c
formulations for the tests considered here, we expect that, when used in
the evolution of binary NSs, the Z4c system would also yield violations
that are of about 1 order of magnitude smaller than those with the
BSSNOK formulation.

\subsection{Fully covariant CCZ4 in black hole spacetimes}
\label{sec:Cvsc} 

As a concluding section, we now discuss how it is possible to
employ a covariant CCZ4 formulation, \ie with $\kappa_3 = 1$, also
for spacetimes containing BH singularities. We recall that in cases where
no singularity is present, as for example in the evolution of a TOV, the
fully covariant CCZ4 system is stable and the standard constraint-damping
prescription leads to results similar to the noncovariant CCZ4
formulation as well as to the Z4c systems (\cf Fig.~\ref{fig:tov_constr}). However, in those cases in which a BH is
present, either initially or when it is formed during the evolution, the
fully covariant CCZ4 system coupled with the constraint-damping scheme
has shown exponentially growing modes (\cf Fig.~4 in paper I).

Even though we could not identify the exact cause of the instability (see
discussion in Sec.~\ref{sec:CCZ4}), it is clear that in
Eqs.~\eqref{eq:dt_K} and \eqref{eq:dt_Theta} the \emph{constant} damping
coefficient $\kappa_1$ is always multiplied by the lapse function
$\alpha$. Because our singularity-avoiding slicing reduces considerably
the value of the lapse near the singularity, it is clear that the
benefits introduced by the damping term $\kappa_1$ are severely
suppressed right there where the violations are the largest. Fortunately,
these considerations suggest two new prescriptions for the damping terms.
In detail, we replace the constant $\kappa_1$ in  
Eqs.~(\ref{K_eq})-(\ref{Gamma_eq}) with one of the following functions:
\begin{align}
\label{eq:new_kappa_1}
\kappa_1 & \rightarrow \frac{\kappa_1}{\alpha}\,,\\
\label{eq:new_kappa_2}
\kappa_1 & \rightarrow \frac{\kappa_1}{2} \left(\alpha + 
\frac{1}{\alpha}\right)\,.
\end{align}
In this way, the product $\alpha\kappa_1$ is not approaching zero anymore
near the singularity. Note that the Z4 damping terms, \ie the terms containing
$\kappa_1$, are not fully covariant anymore, but only spatially
covariant, since we introduced an explicit dependency on the slicing. The
main Z4 evolution equations on the other hand remain unchanged and fully
covariant.

Although the two prescriptions Eqs.~\eqref{eq:new_kappa_1} and
\eqref{eq:new_kappa_2} are slightly different in their local and
asymptotic behavior, they yield very similar results, and we focus on the
form \eqref{eq:new_kappa_1} in the following. To validate the
effectiveness of the new prescription, we recompute the tests performed
in the previous section and compare them with those obtained with the
noncovariant version of the CCZ4 formulation (\ie with $\kappa_3 = 0.5$)
and with the Z4c formulations. Since we use the same computational
infrastructure, numerical methods and gauge conditions, the only
differences between the two systems are the Z4 source terms, which were
truncated in the noncovariant CCZ4 version and are completely removed in
the Z4c system (see Eqs.~\eqref{gamma_eq}-\eqref{Gamma_eq}).

\begin{figure}
  \centering
  \includegraphics[width=0.98\columnwidth]{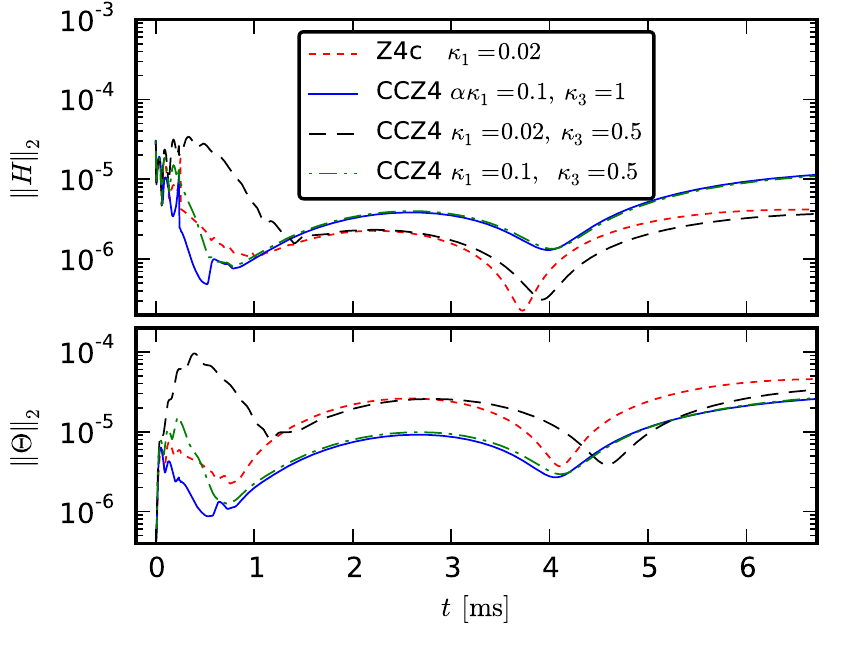}
  \caption{Comparison between the noncovariant and covariant versions 
    of the CCZ4 and the noncovariant Z4c systems in evolutions of an
    unstable TOV star collapsing to a BH. Shown is the time evolution of
    the $L_2$ norm of the Hamiltonian constraint violation (top panel)
    and of the $\Theta$ variable (bottom panel), computed excluding the
    interior of the BH. }
  \label{fig:tovcoll_constr}
\end{figure}

The comparison is shown in Fig.~\ref{fig:sbh_constr}, which presents
results obtained with two values of the damping parameter, namely
$\kappa_1 = 0.1$ and $\kappa_1=0.02$, in evolutions of a single
nonrotating BH. The latter allows a direct comparison with the
noncovariant CCZ4 and Z4c results, and indeed the constraint violations
are very similar in this case. Larger damping values lead to lower values
of $\Theta$ and of the violations of the momentum constraints. Even
though the two damping parameters are significantly different, the
violations of the constraints change by less than 1 order of
magnitude over the timescale of this evolution, \ie $5
\usk\milli\second$. A significant effect, however, is observed in the black
hole mass, which shows deviations of $0.1\%$ in the first case and
$2.8\%$ in the second case (see Table~\ref{tab:bh_mass}).

Overall, the behavior of the covariant CCZ4 constraints matches well the
noncovariant version of the CCZ4 system for the same value of the
damping parameter, for example $\kappa_1=0.1$ in
Fig. \ref{fig:tovcoll_constr}. However, the two versions of the CCZ4
formulations differ in the value of the final BH mass after collapse namely, the covariant one shows a difference of $1.01\%$ with respect to
the initial gravitational mass of the NS, while the noncovariant one
shows a $0.51\%$ error. A study of the influence of different damping
parameters on the values of the BH mass is presented in Table
\ref{tab:bh_mass}. In practice, larger values of the damping lead to more
reliable estimates of the mass; for example, $\kappa_1 = 0.17$ reduces the
error to $0.23\%$.

We believe that the new prescriptions for the damping term Eqs.~\eqref{eq:new_kappa_1} and \eqref{eq:new_kappa_2} are important for two
distinct reasons. First, they allow for the use of a covariant CCZ4
formulation also in singular BH spacetimes. This was a limitation of our
approach in paper I that has been successfully overcome. Second, these
results shed some light on the behavior of the ``nonprincipal part''
constraint-damping terms in the CCZ4 system, although a complete and
closer comparison with the Z4c results presented in Ref.~\cite{Hilditch2012}
cannot be performed because of the different boundary conditions
employed.

\section{Conclusions}
\label{sec:summary}

We have compared numerically the performance of several conformal and
traceless formulations of the Z4 system with the BSSNOK formulation in
simulations of spacetimes with and without matter, in terms of the
suppression of constraint violations and their impact on the accuracy of
physical quantities. We successfully coupled the CCZ4 system presented in
paper I to matter sources, and also found that the fully covariant CCZ4
version leads to stable evolutions for the case of NSs, while it develops
instabilities as soon as a BH is present. We created a modified CCZ4
version that completely eliminates those instabilities for BH
spacetimes also, and which is still covariant. In addition, we implemented
the Z4c formulation described in Ref.~\cite{Hilditch2012} in our numerical
framework and compared its properties to those of the CCZ4 formulation.

We have found that the noncovariant CCZ4 formulation introduced in
paper~I is stable when coupled to matter sources in simulations of stable
and unstable spherical NSs, as well as of merging binary NSs. In
comparison to the BSSNOK system, the CCZ4 formulation reduces the
constraint violations by 1 order of magnitude for simulations of binary
NSs and by 2 orders for single NSs. We have also demonstrated the
convergence of our implementation for simulations of stable NSs, as well
as those of binary NS mergers, with a convergence order in the range
$1.2$-$2$, mainly limited by the hydrodynamic evolution scheme. By
comparing the CCZ4 and the BSSNOK evolutions, and by using the large
differences in the constraint violations, we could estimate the influence
of the latter on physical quantities. In particular, we could demonstrate
that the impact of the constraint violations on the GW signal from the
binary NS mergers is smaller than or comparable to the numerical errors.

Furthermore, a comparison of the different conformal Z4 versions, namely
the noncovariant and covariant CCZ4 formulations, and the noncovariant
Z4c, has shown that they have an almost identical behavior in terms of
constraint violations as long as no singularity is present (\eg stable
TOV). In evolutions of BH spacetimes, or when a BH is produced as a result
of a collapse, we find that values of the constraint-damping parameter
larger than the ones proposed for the Z4c formulation (\ie $\kappa_1 =
0.1$ in place of $\kappa_1 = 0.02$) lead to lower constraint violations
and do not produce late-time instabilities. Finally, we have also found
that the modification of the damping terms mentioned above is useful not
only for a stable evolution of the covariant CCZ4 system, but 
also for reducing the drift in the BH mass for both CCZ4 versions.

Overall, we recommend CCZ4 as the standard formulation of the Einstein
equations to be used in numerical relativity evolutions. In cases where
constraint violations are problematic, \eg when using
constraint-violating initial data, the CCZ4 formulation is clearly
superior. On the other hand, when using constraint-satisfying initial
data, the reduction of constraint violations is accompanied by errors
that are very similar to those obtained with the BSSNOK formulation. No
additional computational costs are needed, and simple Sommerfeld radiative
boundary conditions are sufficient to obtain stable evolutions. On the
basis of the tests where we performed a direct comparison, the
performance of Z4c and CCZ4 seems to be comparable. However, CCZ4 has the
advantage of being covariant, while Z4c has been successfully tested with
constraint-preserving boundary conditions; see Ref.~\cite{Hilditch2012}. In
simulations of binary neutron star mergers, a main limitation of the
accuracy is given by the hydrodynamic part. For this application, using
higher-order schemes for the hydrodynamic equations, such as those
presented in Ref.~\cite{Radice2013b}, might have a larger impact than the
choice between CCZ4 and Z4c.


\begin{acknowledgments}

It is a pleasure to acknowledge F. Galeazzi, D. Radice and K. Takami for
numerous useful discussions. We are grateful to Carles Bona and Carlos
Palenzuela for important suggestions. Partial support comes from the DFG grant
SFB/Transregio 7, and from ``CompStar'', a Research Networking Programme of
the European Science Foundation. The calculations were performed on the
SuperMUC cluster at the LRZ and on the Datura cluster at the AEI.

\end{acknowledgments}


\end{document}